\documentclass[10pt, twocolumn, conference]{IEEEtran}    

\IEEEoverridecommandlockouts                              
\title{Distortion Exponent in MIMO Channels\\ with Feedback}

\author{%
  \authorblockN{Deniz G\"{u}nd\"{u}z\authorrefmark{1}\authorrefmark{2},
    Andrea Goldsmith\authorrefmark{2}, H. Vincent Poor\authorrefmark{1}
  }\\
  \authorblockA{%
    \authorrefmark{1}Department of Electrical Engineering, Princeton University, Princeton, NJ.\\
  }
  \authorblockA{%
    \authorrefmark{2}Department of Electrical Engineering, Stanford University, Stanford, CA.\\
  }
  Email: dgunduz@princeton.edu, andrea@wsl.stanford.edu, poor@princeton.edu
  \thanks{This research was supported by the National Science Foundation under Grant CNS-06-25637 and by the Defense Threat Reduction Agency
under Grant HDTRA1-08-1-0010.}
}
\date{November, 2007}
\usepackage{amsfonts}
\usepackage{amssymb}
\usepackage{amsmath}
\usepackage{amscd}
\usepackage{psfrag, graphicx}

\newtheorem{thm}{Theorem}[section]

\newtheorem{rem}{Remark}[section]

\begin{document}
\maketitle
\thispagestyle{empty}
\pagestyle{empty}

\begin{abstract}
The transmission of a Gaussian source over a block-fading multiple antenna channel in the presence of a feedback link is considered. The feedback link is assumed to be an error and delay free link of capacity $1$ bit per channel use. Under the short-term power constraint, the optimal exponential behavior of the end-to-end average distortion is characterized for all source-channel bandwidth ratios. It is shown that the optimal transmission strategy is successive refinement source coding followed by progressive transmission over the channel, in which the channel block is allocated dynamically among the layers based on the channel state using the feedback link as an instantaneous automatic repeat request (ARQ) signal.
\end{abstract}

\section{Introduction}\label{s:intro}

High quality transmission of multimedia signals over wireless channels is an important research problem from both theoretical and practical aspects. While the rich content of multimedia signals requires transmission at high data rates, the unreliable nature of the wireless medium due to fading makes it hard to sustain these high rates at all times. An appropriate performance measure for these applications is the average end-to-end signal distortion, which will be caused by both the compression of the continuous amplitude source samples and the errors over the channel.

Here, we consider the transmission of a discrete-time Gaussian source over a block fading multiple-input multiple-output (MIMO) channel. We assume that the instantaneous channel state information (CSI) is available at the receiver. Motivated by real-time applications, we consider a strict delay constraint such that $K$ source samples are to be transmitted
within a single channel block of $N$ channel uses. We
define the \textit{bandwidth ratio} of the system as $\mathit{b}=\frac{N}{K}$ channel uses per source sample, and analyze the system performance with respect to $\textit{b}$.


We are interested in the high $SNR$ behavior of the
average distortion, $D(SNR)$, which is characterized by the
\emph{distortion exponent} \cite{Emin}:
\begin{equation}\label{d:dist_exp}
\Delta=-\lim_{SNR\rightarrow\infty}\frac{\log D(SNR)}{\log SNR}.
\end{equation}

In addition to the direct channel, we assume that there is a feedback link available from the receiver to the transmitter, which can transmit $1$ bit per channel use (bpcu) without error or delay. The effect of feedback on the performance of MIMO channels has been the subject of considerable research. In the block fading scenario that we consider, the feedback link of a finite number of bits per channel block can be used to transmit the channel state information to the transmitter \cite{Kim:IT:07}, to ask for additional parity bits in an automatic repeat request (ARQ) system \cite{ElGamal:IT:06}, or to delay the transmission until channel conditions improve \cite{Sharif:ISIT:07}. In our model, we strictly constrain the transmission to a single block, and power allocation among blocks is not possible due to the short-term power constraint.

Our main result is a characterization of the distortion exponent for all bandwidth ratios and an arbitrary number of transmit and receive antennas in the presence of $1$ bpcu feedback. In the optimal transmission scheme, source samples are compressed into $n$ successive refinement source layers. These layers are channel coded at different rates to provide unequal error protection against channel fading, and are transmitted progressively over the channel. The feedback link is used to inform the transmitter about the success of decoding of the source layer that is being transmitted. As soon as an ACK signal is received from the receiver, the transmitter starts transmitting the channel codeword corresponding to the next source layer. This transmission strategy can be considered as an instantaneous ARQ scheme. ARQ feedback is widely used in wireless networks to improve the reliability of the system; however, it is an important open research problem whether the ARQ scheme is the best way to make use of the finite rate feedback resources. The answer, in general, depends on the performance measure and the operating regime. Here, under an extreme model of instantaneous ARQ feedback, we prove its optimality in terms of the end-to-end distortion exponent.



The characterization of the distortion exponent for fading channels has recently been investigated in several studies. The distortion exponent is first defined in \cite{Emin}. The achievable distortion exponent in the case of MIMO channels is studied in \cite{Gunduz:IT:08}-\cite{Bhattad:IT:08}. It is shown in \cite{Gunduz:IT:08} that, by using a successive refinement source encoder and superposition coding over the channel \cite{Shamai:IT:03}, the optimal distortion exponent can be achieved in systems with one degree-of-freedom at all bandwidth ratios. However, the optimality of this scheme is limited to high bandwidth ratios in the case of general MIMO systems. While Bhattad et al. \cite{Bhattad:IT:08} extended the optimality of this scheme to a larger set of bandwidth ratios using a different power allocation strategy, the characterization of the distortion exponent at all bandwidth ratios remains open. These results show that, at least in certain scenarios, successive refinement source coding concatenated with superposition channel coding can adapt to the channel fading optimally as it achieves the distortion exponent upper bound obtained by assuming perfect channel state information at the transmitter. The transmission algorithms proposed in \cite{Gunduz:IT:08} are studied in the finite $SNR$ regime in \cite{Etemadi:ISIT:06}-\cite{Tian:IT:08}.

Our result in this paper shows that the availability of the 1 bpcu feedback link provides adaptation to the channel state independent of the number of degrees of freedom of the system. We note here that the feedback link does not improve the diversity multiplexing tradeoff (DMT) performance under the short-term power constraint since the DMT analysis does not allow rate adaptation. Source transmission over a block-fading MIMO channel with feedback is also studied in \cite{Kim:SP:09}, in which the authors consider finite bit feedback in the form of quantized channel state information. A hybrid digital-analog coding scheme is proposed in \cite{Kim:SP:09}, which uses the feedback information to assign the rate of the digital component, and improves the distortion exponent even with a feedback link of a single bit per channel block. 

The rest of the paper is organized as follows. In Section \ref{s:system} we introduce the system model in mathematical terms. The main result of the paper is presented in Section \ref{s:result}, and its proof is given in Section \ref{s:proof}. Section \ref{s:conc} concludes the paper.

\section{System Model}\label{s:system}

We consider an independent identically distributed (i.i.d.) complex Gaussian source $\{s_k\}_{k=1}^\infty$ with independent zero-mean real and imaginary
components each with variance $1/2$ (i.e., $\mathcal{CN}(0,1)$). For the analysis, we assume compression strategies that meet the optimal distortion-rate function given by $D(R)=2^{-R}$ where $R$ is the source coding rate in bits per source sample. Due to the delay requirement of the application, $K$ source samples are to be transmitted within a single channel block of $N$ channel uses, which corresponds to a bandwidth ratio of $\textit{b}=N/K$ channel uses per source sample. We want to characterize the high $SNR$ behavior of the end-to-end distortion with respect to the bandwidth ratio $\mathit{b} \geq 0$.

The underlying communication channel is modeled as a slow-fading MIMO channel with $M_t$ transmit and $M_r$ receive antennas. The channel model is
\begin{equation}\label{channel}
\mathbf{y}[i]=\sqrt{\frac{SNR}{M_t}}\mathbf{H}\mathbf{x}[i]+\mathbf{z}[i],
\hspace{.3in} i=1,\ldots,N
\end{equation}
where $\sqrt{\frac{SNR}{M_t}}\mathbf{x}[i]$ is the transmitted signal at time $i$,
$\mathbf{Z}=[\mathbf{z}_1,\ldots,\mathbf{z}_N]\in\mathbb{C}^{M_r\times N}$
is complex Gaussian noise with i.i.d $\mathcal{CN}(0,1)$ entries, and $\mathbf{H} \in \mathbb{C}^{M_r\times M_t}$ is the channel matrix, which has i.i.d. entries
with $\mathcal{CN}(0,1)$. The realization of
the channel matrix $\mathbf{H}$ is assumed to be known by the
receiver and unknown by the transmitter, while the transmitter knows
the statistics of $\mathbf{H}$. The codeword, $\mathbf{X}=\bigg[
\mathbf{x}[1], \ldots, \mathbf{x}_L \bigg]\in\mathbb{C}^{M_t\times
N}$ is normalized so that it satisfies
$tr(E[\mathbf{X}^\dag\mathbf{X}])\leq M_tN$.  We define
$M_*=\min\{M_t,M_r\}$ and $M^*=\max\{M_t,M_r\}$.

In addition to the direct channel in (\ref{channel}), we also assume that a zero-delay error-free feedback link of capacity $1$ bpcu is available from the receiver to the transmitter. We denote the feedback signal at time instant $i$ by $v[i]$, where $v[i] \in \{0,1\}$ for $i=1,\ldots,N$. The feedback symbol $v[i]$ can depend on the channel matrix $\mathbf{H}$, which is known perfectly at the receiver, and the received signals until time $i$, i.e.,
\begin{equation*}
    v[i] = g_i(H, y[1], \ldots, y[i-1]),
\end{equation*}
where $g_i$ is the feedback encoder at time $i$. The channel input $\mathbf{x}[i]$ of the transmitter is given by
\begin{equation*}
    \mathbf{x}[i] = f_i(\mathbf{s}^K, v[1], \ldots, v[i-1]),
\end{equation*}
where $f_i$ is the encoding function of the transmitter at time instant $i$ and $\mathbf{s}^K=[s_1,\ldots, s_K]$ is the source block.

The decoder maps the received signal
$\mathbf{Y}=\bigg[\mathbf{y}[1],\ldots,\mathbf{y}[N]\bigg]\in\mathbb{C}^{M_r\times
N}$ to an estimate $\hat{\mathbf{s}}^K \in\mathbb{C}^K$ of the source.
The average distortion $D(SNR)$ is defined as $\frac{1}{K}E[|\mathbf{s}^K-\hat{\mathbf{s}}^K|^2]$, where the expectation is with
respect to the source, channel and noise distributions.

We are interested in
the high $SNR$ behavior of the expected distortion. We optimize the
system performance to maximize the \textit{distortion exponent}
defined in (\ref{d:dist_exp}). A distortion exponent of $\Delta$
means that the expected distortion decays as $SNR^{-\Delta}$ with increasing $SNR$.

In order to obtain the end-to-end average distortion for our proposed transmission
strategy, we need to characterize the error rate of the MIMO
channel. Since we are interested in the high $SNR$ regime, we use the
outage probability, which has the same exponential
behavior as the channel error probability with long enough codewords \cite{Tse}. For a family
of codes $\mathfrak{C}(SNR)$ at rates $R(SNR)$, the multiplexing
gain is defined as $r\triangleq \lim_{SNR\rightarrow \infty} \frac{R(SNR)}{\log SNR}$, and the diversity advantage is defined as $d(r) \triangleq - \lim_{SNR\rightarrow\infty}\frac{\log P_{out}(SNR)}{\log SNR}$, where $P_{out}(SNR)$ is the outage probability of the code. The diversity gain $d^*(r)$ is defined as the supremum of the diversity advantage over all possible code families with multiplexing gain $r$. The diversity-multiplexing gain tradeoff (DMT) is characterized as follows \cite{Tse}.

\begin{thm} \label{divmimo}
For an $M_t\times M_r$ MIMO block fading channel, the optimal tradeoff curve $d^*(r)$ is given by a piecewise-linear function
connecting the points $(k,d^*(k))$, $k=0,1,\ldots,M_*$, with $d^*(k) = (M_t-k)(M_r-k)$.
\end{thm}

\section{Distortion Exponent with Feedback}\label{s:result}

In this section we present our main result in which we characterize the optimal distortion exponent of the MIMO system with feedback. We first propose an upper bound on the achievable distortion exponent of this system.

\begin{thm}\label{t:upperbound}{}
In transmission of the i.i.d. zero mean unit variance complex Gaussian source over a slow fading $M_t \times M_r$ MIMO
channel in the presence of a delay and error free feedback link of $1$ bpcu, the distortion exponent is upper bounded by
\vspace{-.15in}
\begin{equation}\label{e:dist_exp}
\sum_{i=1}^{M_*}\min\left\{\textit{b}, 2i-1+ M^*- M_* \right\}.\vspace{-.06in}
\end{equation}
\end{thm}

\begin{proof}
This is the same upper bound proposed in \cite{Gunduz:IT:08} and \cite{Caire:IT:07} when there is no feedback link. This upper bound is obtained by assuming that the CSI at each channel block is provided to the transmitter by a genie. Note that feedback does not increase the point-to-point instantaneous capacity, and power allocation among channel blocks is not allowed due to the short-term power constraint. Hence, the same distortion exponent upper bound is obtained with or without the feedback link when CSI is assumed at the transmitter.
\end{proof}

The next theorem states that the above upper bound can be achieved in any given $M_t \times M_r$ MIMO system in the presence of a $1$-bit delay- and error-free feedback link.

\begin{thm}\label{t:fdbk_ach}
In transmission of an i.i.d. complex Gaussian source with distribution $\mathcal{CN}(0,1)$ over a slow fading $M_t \times M_r$ MIMO
channel with $1$ bpcu feedback link, the optimal distortion exponent is characterized by (\ref{e:dist_exp}).
\end{thm}

\begin{figure}
\centering
\includegraphics[width=3.0in]{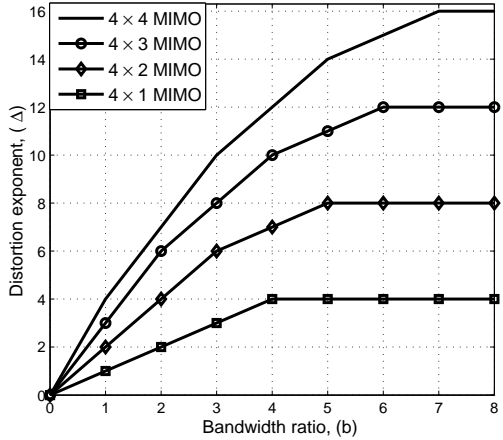}
\caption{Distortion exponent of various MIMO systems in the presence of a noise-free zero-delay feedback link of capacity $1$ bpcu.} \label{f:DEf}
\end{figure}

We illustrate the optimal distortion exponent of a MIMO system with $M_t=4$ transmit antennas and $M_r=1,2,3$ and $4$ antennas in Fig. \ref{f:DEf}. Note that these distortion exponent values can be achieved without the feedback link only for $0 \leq b \leq (M^*-M_*+1)/M_*$ and $b \geq M_tM_r$ using the superposition of infinite source layers \cite{Gunduz:IT:08}, \cite{Bhattad:IT:08}.

\begin{rem}
The achievability scheme in the proof of Theorem \ref{t:fdbk_ach} in Section \ref{s:proof} uses successive refinement source coding followed by progressive transmission of the layers over the channel. Each channel codeword is transmitted until an ACK signal is received from the receiver over the feedback link. In the absence of a feedback link, progressive transmission of layers was considered in \cite{Gunduz:IT:08} with optimal but fixed allocation of the channel block among the layers. This falls short of the upper bound since fixed channel allocation loses in terms of the multiplexing gain and cannot exploit the good states of the channel as well as the proposed scheme with a feedback link.
\end{rem}

\begin{rem}
The distortion exponent upper bound in Theorem \ref{t:upperbound} can be achieved in MISO/SIMO systems without the feedback link \cite{Gunduz:IT:08} and in general MIMO systems for certain bandwidth ratios \cite{Gunduz:IT:08}, \cite{Bhattad:IT:08}. However, the optimal transmission scheme requires superposition coding and successive decoding. We show here that a simple point-to-point transmission scheme can achieve this performance in the presence of a feedback link of $1$ bpcu.
\end{rem}

\section{Proof of Theorem \ref{t:fdbk_ach}} \label{s:proof}
\begin{proof}
We use the following transmission and feedback coding scheme: The encoder compresses the $K$ source samples into $n$ layers using successive refinement source coding. These compressed layers are than transmitted over the channel in a progressive manner. Let $R_1, \ldots, R_n$ be the channel transmission rates of these source layers. Due to the fixed bandwidth ratio $b$, this imposes a source coding rate of $bR_k$ for layer $k$.  While the channel and source coding rates are fixed for each layer, the transmission time is dynamic based on the feedback signal from the receiver. The transmitter starts transmitting the channel codeword of the first layer. At each time instant $i$, the feedback link is set to $0$ if the current transmitted layer has not been decoded at the receiver yet. It is set to $1$ when the receiver decodes the current layer. Note that this can be considered as an instantaneous ARQ scheme.

We assume Gaussian codebooks at the transmitter with identity input covariance matrix. We denote the instantaneous achievable capacity of the channel with channel realization $\mathbf{H}$ as $C(\mathbf{H})$. We have
\vspace{-.05in}
\begin{equation}
        C(\mathbf{H}) \triangleq \log \det \left(\mathbf{I} + \frac{SNR}{M_t} \mathbf{H H}^\dag \right),\vspace{-.05in}
\end{equation}
where $^\dag$ denotes the complex conjugate operation.

In the proposed transmission scheme, a layer is in outage if the receiver cannot successfully decode this layer before the end of the channel block. Once the layer $k$ is in outage, none of the remaining layers can be decoded at the receiver as the transmitter has no time to transmit these layers. However, note that even if the remaining layers were available at the receiver, these layers are useless in the absence of layer $k$ due to the properties of successive refinement source coding.

In order to obtain a vanishing average distortion at the receiver with increasing $SNR$, we scale the transmission rate of each source layer with $SNR$ as $R_k= r_k \log SNR$, where $r_k$ is the multiplexing gain of layer $k$. For $k=0, 1, \ldots, n$, let $\mathcal{A}_k$ denote the event that exactly $k$ source layers can be decoded at the receiver. We have $Pr\{ \mathcal{A}_0\} =  Pr\{ r_1 \log SNR > C(\mathbf{H}) \}$. For the event $\mathcal{A}_k$, all layers up to layer $k$ need to be decoded at the destination while there is not sufficient time to decode layer $k+1$. We have
\begin{align}
    Pr\{&   \mathcal{A}_k\}  =Pr\left\{ \frac{R_1}{C(\mathbf{H})} + \cdots + \frac{R_k}{C(\mathbf{H})} \leq 1   \right. \nonumber \\
     & ~~~~~~~~~~~~~~~~~~~~~~~~~~~~~~ \left. < \frac{R_1}{C(\mathbf{H})} + \cdots + \frac{R_{k+1}}{C(\mathbf{H})} \right\}, \nonumber \\
      &= Pr\{ (r_1+ \cdots+ r_k) \log SNR \leq C(\mathbf{H}) \nonumber \\
     & ~~~~~~~~~~~~~~~~~~~~~~~~~ < (r_1+ \cdots+ r_{k+1}) \log SNR  \}, \nonumber
     \end{align}
     \begin{align}
      = Pr\{ \bar{r}_k \log SNR \leq C(\mathbf{H}) < \bar{r}_{k+1} \log SNR  \}, \label{e:eventAk}
\end{align}
where we define $\bar{r}_k \triangleq \sum_{j=1}^k r_j$. Finally, for the last layer we have $Pr\{ \mathcal{A}_n\} =  Pr\{ \bar{r}_n \log SNR \leq C(\mathbf{H}) \}$.

Let $P_{out}(r) \triangleq Pr\{ r\log SNR > C(\mathbf{H})\}$ be the outage probability at multiplexing gain $r$. (We suppressed the dependence on the channel realization and $SNR$ for simplicity of notation.) For $k=2,\dots,n-1$ we can write (\ref{e:eventAk}) as
\begin{align}
    Pr\{ \mathcal{A}_k \} = & P_{out}(\bar{r}_{k+1}) - P_{out}(\bar{r}_k),  \label{e:eventout} \\
        \doteq & SNR^{-d^*(\bar{r}_{k+1})} - SNR^{-d^*(\bar{r}_k)} \\
        \doteq & SNR^{-d^*(\bar{r}_{k+1})}.
\end{align}
We have $Pr\{ \mathcal{A}_n\} \doteq 1- SNR^{-d^*(\bar{r}_n)} $.

\psfrag{r}{$r$}\psfrag{dr}{$d^*(r)$}\psfrag{DMT}{DMT curve}\psfrag{b}{$\scriptstyle b$}
\psfrag{d1}{$\hspace{-.1in} \scriptstyle\Delta = d^*(\bar{r}_1)$}\psfrag{d2}{\hspace{-.1in} $\scriptstyle d^*(\bar{r}_2)$} \psfrag{d3}{\hspace{-.1in} $\scriptstyle d^*(\bar{r}_3)$}
\psfrag{d4}{\hspace{-.1in} $\scriptstyle d^*(\bar{r}_4)$} \psfrag{d5}{\hspace{-.1in} $\scriptstyle d^*(\bar{r}_5)$} \psfrag{d6}{\hspace{-.1in} $\scriptstyle d^*(\bar{r}_6)$}
\psfrag{br1}{$\scriptstyle\bar{r}_1$}\psfrag{br2}{$\scriptstyle\bar{r}_2$}\psfrag{br3}{$\scriptstyle\bar{r}_3$}
\psfrag{br4}{$\scriptstyle\bar{r}_4$}\psfrag{br5}{$\scriptstyle\bar{r}_5$}\psfrag{br6}{$\scriptstyle\bar{r}_6$}
\psfrag{bbr1}{$\scriptstyle br_1$}\psfrag{bbr2}{$\scriptstyle br_2$}\psfrag{bbr3}{$\scriptstyle br_3$}
\psfrag{bbr4}{$\scriptstyle br_4$}\psfrag{bbr5}{$\scriptstyle br_5$}\psfrag{bbr6}{$\scriptstyle br_6$}
\begin{figure}
\centering
\includegraphics[width=2.3in]{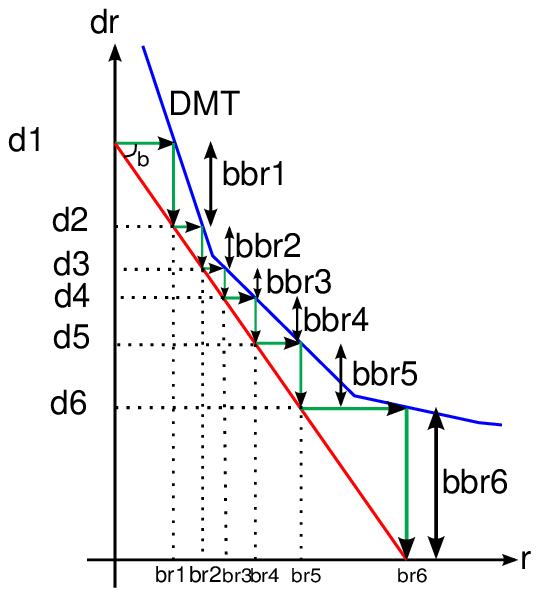}
\caption{Illustration of the Equations (\ref{e:dexp_f1})-(\ref{e:dexp_f4}) for six layers on the DMT curve.} \label{f:fdbk}
\end{figure}

The distortion achieved at the receiver upon successfully decoding $k$ layers, $k=0,1,\ldots,n$, is given by
\begin{align}
    D_k & \triangleq D \bigg(b(R_1+\cdots+ R_k) \bigg) \nonumber \\
        & = 2^{-b(R_1+\cdots +R_k)} \nonumber \\
        & = SNR^{-b \bar{r}_k}
\end{align}
where we define $D_0 \triangleq 1$. Then we can write the average distortion expression as
\begin{align}
    E[D] = & \sum_{k=0}^n Pr\{A_k\} D_k, \\
        \doteq & \sum_{k=0}^n SNR^{-d^*(\bar{r}_{k+1})} SNR^{-b \bar{r}_k}, \\
        \doteq & SNR^{- \min_{0\leq k \leq n} \{d^*(\bar{r}_{k+1}) + b \bar{r}_k \} },
\end{align}
where we define $d^*(\bar{r}_{n+1})=0$ and $\bar{r}_0=0$. Hence, for $n$ layers, the distortion exponent maximization problem can be written as

\begin{eqnarray}\label{m:dexp_opt}
\Delta_n^{f} &=& \max _{r_1,\ldots,r_n} \min_{0\leq k\leq n}\left\{ d^*(\bar{r}_{k+1}) + b \bar{r}_k \right\} \\
&\mbox{s.t.}& 0\leq r_k \mbox{ for } k=1,\ldots,n. \nonumber
\end{eqnarray}

The optimal distortion exponent is achieved by having all the exponents equal, that is,
\begin{eqnarray}
b r_n &=& d^*(\bar{r}_n),  \label{e:dexp_f1} \\
d^*(\bar{r}_n) + b r_{n-1} &=& d^*(\bar{r}_{n-1}), \label{e:dexp_f2} \\
& \dots & \nonumber \\
d^*(\bar{r}_3)+b r_2 &=&d^*(\bar{r}_2), \label{e:dexp_f3} \\
d^*(\bar{r}_2)+b r_1 &=&d^*(\bar{r}_1),\label{e:dexp_f4}
\end{eqnarray}
where the corresponding distortion exponent is $\Delta^f_n = d^*(\bar{r}_1)$.

\psfrag{r}{$r$}\psfrag{dr}{$d^*(r)$}\psfrag{DMT}{DMT curve}\psfrag{b}{$b$}
\psfrag{slope}{slope}\psfrag{s}{$\scriptstyle -(M_t+M_r)+2k-1$}
\psfrag{k}{$\scriptstyle k$}\psfrag{k2}{$\scriptstyle k+1$}
\psfrag{M1}{\hspace{-.1in} $\scriptstyle d^*(k)$}\psfrag{M2}{\hspace{-.1in} $\scriptstyle d^*(k+1)$}
\psfrag{bi}{\hspace{-.1in} $b\nearrow$}
\begin{figure}
\centering
\includegraphics[width=2.0in]{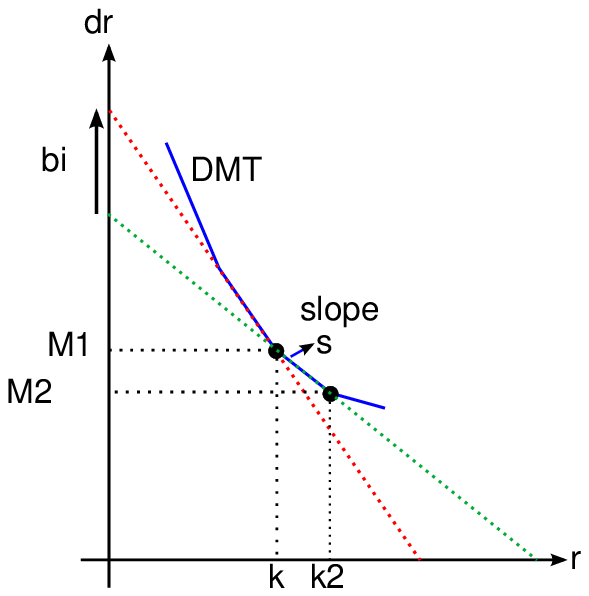}
\caption{Illustration of possible straight lines that lie under the DMT curve with the highest $y$ intercept for different slopes.} \label{f:slope}
\end{figure}

We solve this optimization problem by reducing it to an analytical geometry problem. We first make the following observation: For a finite number of layers $n$, if we mark the solutions of (\ref{e:dexp_f1})-(\ref{e:dexp_f4}) on the DMT curve, the points $(\bar{r}_k, d^*(\bar{r}_{k+1}))$, for $k=0,\ldots, n$, lie on a straight line with slope $-b$. (See Fig. \ref{f:fdbk}.) We call this line the $\Delta$-line. Since the $y$-intercept of the $\Delta$-line, that is, $d^*(r_1)$, is the corresponding distortion exponent, we would like a $\Delta$-line with the highest possible $y$-intercept. Note that since the DMT curve is piece-wise linear, the $\Delta$-line either touches the DMT curve at one of the corner points, or coincides with one of the pieces of the DMT curve (when its slope $-b$ is equal to the slope of one of the pieces).

In the remaining part of the proof, we first identify the highest possible point at which a $\Delta$-line can intersect the $y$-axis. Then we will show that, in the limit of infinitely many layers, it is possible to assign multiplexing gains so that the achieved distortion exponent will be characterized by the $y$-coordinate of that intersection point.

First assume that any $\Delta$-line corresponds to a multiplexing gain assignment that satisfies (\ref{e:dexp_f1})-(\ref{e:dexp_f4}). Hence, the $y$-intercept of the $\Delta$-line gives the corresponding distortion exponent. As shown in Fig. \ref{f:slope}, with increasing $b$, the highest possible $y$-intercept of the possible $\Delta$-lines increases (up to $M_tM_r$ which is the $y$-intercept of the DMT curve). For $b \geq M_t+M_r-1$, which is the slope of the first piece of the DMT curve intersecting the $y$-axis, the $\Delta$-line passing through $(0, M_tM_r)$ will lie under the DMT curve. This corresponds to the maximum possible distortion exponent of $M_tM_r$ for all $b \geq M_t+M_r-1$. If $b=(M_t+M_r)-(2k+1)$, the highest possible $y$-intercept is $[(M_t+M_r)-(2k+1)](k+1) + d^*(k+1)$ as seen from Fig. \ref{f:slope}. In between the points $((M_t+M_r)-(2k+1), [(M_t+M_r)-(2k+1)](k+1) + d^*(k+1))$ the corresponding distortion exponent is a linear function of $b$. That is, we obtain a piecewise linear function of the bandwidth ratio $b$. It is not difficult to prove that the highest possible $y$-intercept of a $\Delta$-line with slope $-b$ is characterized by the function in (\ref{e:dist_exp}).

\begin{figure}
\centering
\includegraphics[width=2.5in]{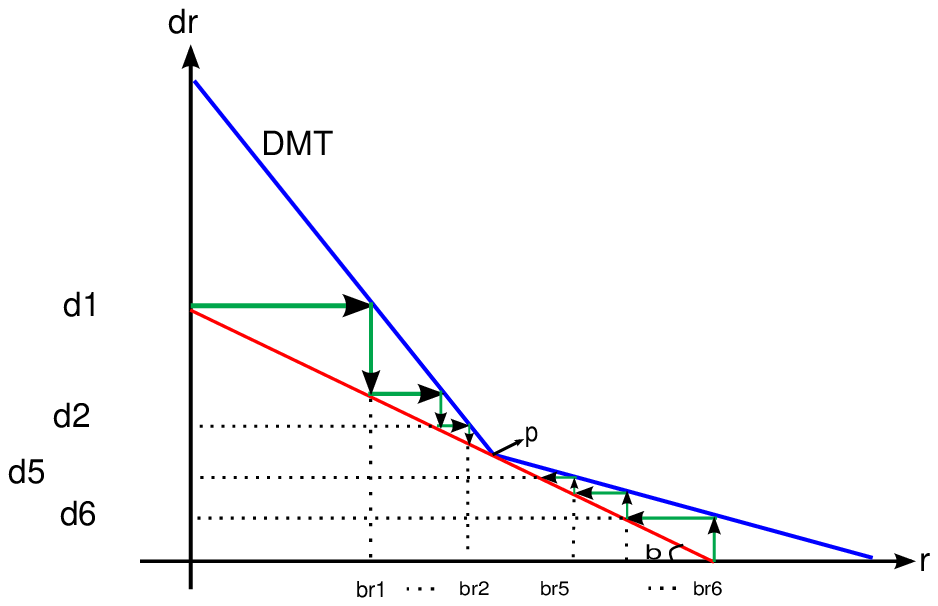}
\caption{Illustration of the multiplexing gains for a given DMT curve and bandwidth ratio $b$.} \label{f:rates}
\end{figure}

Next, for any given $\Delta$-line, we characterize a multiplexing gain assignment for the layers. We illustrate this process in Fig. \ref{f:rates}. Let $n=2m$ be the number of source layers. For any given DMT curve and bandwidth ratio $b$, we identify the $\Delta$-line with the highest $y$-intercept. Assume that this line touches the DMT curve at $(j,d^*(j))$ for some $j=1,\ldots,M_*$. We let $d^*(r_1)$ be the $y$-intercept of this straight line, which also determines $r_1$ uniquely. We fix $d^*(\bar{r}_2)=d^*(\bar{r}_1)-b\bar{r}_1$, which also determines $\bar{r}_2$, and hence $r_2$. We continue this process $m$ steps until the multiplexing gain $r_m$ is determined. (See Fig. \ref{f:rates}.) Note that, due to the monotone decreasing nature of the DMT curve we have $\bar{r}_1 < \bar{r}_2 < \cdots < \bar{r}_m$. Similarly, we fix the $x$-intercept of the chosen $\Delta$-line as $\bar{r}_n$, which gives us $d^*(\bar{r}_n)$. We find $r_n$ such that $br_n= d^*(\bar{r}_n)$, which fixes $\bar{r}_{n-1}$. Similarly, we find $r_{n-1}$ such that $br_{n-1} = d^*(r_{n-1}) - d^*(r_n)$. Continuing $m$ steps, we identify the multiplexing gains $r_{m+1}, \ldots, r_n$. From (\ref{m:dexp_opt}), we have
\begin{eqnarray}\label{m:dexp_ach}
\Delta_n^{f} &= & \min_{0\leq k\leq n}\left\{ d^*(\bar{r}_{k+1}) + b \bar{r}_k \right\} \\
    &= & d^*(\bar{r}_{m+1}) + b\bar{r}_m. \label{m:dexp_ach2}
\end{eqnarray}
As $n \rightarrow \infty$, we have $\bar{r}_m, \bar{r}_{m+1} \rightarrow j$, hence $d^*(\bar{r}_{m+1})+ b\bar{r}_m \rightarrow d^*(j)+bj = d^*(r_1)$.

If the chosen $\Delta$-line does not touch the DMT curve, we can always move to the right until the two touch each other, which can only lead to a higher $y$-intercept. On the other hand, the slope $-b$ might also be equal to the slope of one of the pieces of the DMT curve, so that the straight line will intersect the DMT curve within that piece. In this case, we can consider a $\Delta$-line with slope $-b+\epsilon$ for some $\epsilon>0$. The multiplexing gain assignment is done using this slope, and the achievable distortion exponent will be at least as high as the $y$-intercept of the straight line with slope $-b+\epsilon$. Letting $\epsilon \rightarrow 0$, we can obtain distortion exponents arbitrarily close to the one given in the theorem.

\end{proof}

\psfrag{r}{$r$}\psfrag{dr}{$d^*(r)$}\psfrag{DMT}{DMT curve}\psfrag{b}{$\scriptstyle  b$}
\psfrag{d1}{$\scriptstyle d^*(\bar{r}_1)$}\psfrag{d2}{$\scriptstyle d^*(\bar{r}_m)$}
\psfrag{d5}{$\scriptstyle d^*(\bar{r}_{m+1})$}\psfrag{d6}{$\scriptstyle d^*(\bar{r}_n)$}
\psfrag{br1}{$\scriptstyle \bar{r}_1$}\psfrag{br2}{$\scriptstyle\bar{r}_m$}
\psfrag{br5}{$\scriptstyle\bar{r}_{m+1}$}\psfrag{br6}{$\scriptstyle\bar{r}_n$}
\psfrag{p}{$(j,d^*(j))$}

\section{Conclusions}\label{s:conc}

We have considered the problem of transmitting a complex Gaussian source over a block fading channel in the high $SNR$ regime. Assuming a zero-delay error-free feedback link of $1$ bpcu, we have characterized the optimal distortion exponent for an arbitrary MIMO system at any given bandwidth ratio. Our result shows that the single bit feedback link, used as an instantaneous ARQ signal, allows the system to adapt to the channel fading optimally in terms of the distortion exponent, since the performance meets the upper bound obtained by assuming perfect channel state information at the transmitter.

Our result sheds some light on the optimal use of finite rate feedback resources. While using the ARQ bits for retransmissions to increase the reliability of a message is a common practice in wireless networks, our result points to an alternative usage of the ARQ feedback, in which the received signal quality dynamically adapts to the channel state by transmitting additional source layers based on the feedback link.

\end{document}